\documentclass[aps,prl,twocolumn,showpacs,preprintnumbers,amsmath,floatfix]{revtex4}

\usepackage{dcolumn}
\usepackage{bm}

\begin{document}
\title{A New Scenario of Confinement and Hadron Spectra}
\author{Ying Chen}
\affiliation{%
\centerline{Institute of High Energy Physics, Chinese Academy of Sciences, 
Beijing 100039, P. R. China}
}
\date{Oct. 17, 2003}
\begin{abstract}
This work presents a new phenomenological description of QCD vacuum 
where the color background field is depicted by a constant parallel vector 
with amplitude proportional to $\sigma\sim 0.28 GeV^2$. The familiar Regge 
relation can be derived directly in a classic meaning. 
A new mass formula of hadrons are conjectured to give a very consistent
description of hadron mass spectra. The lower bounds of the baryon mass 
are determined to be $\sqrt{3\sigma}$ for nucleon and $\sqrt{5\sigma}$ 
for isospin-$\frac{3}{2}$ baryon. In the meanwhile, the mass-square 
differences of 
$1 {}^3 S_1$ and $1 {}^1 S_0$ meson states are obtained to be $2 \sigma$. 
We also predict the mass of the lowest four-quark state to be 
$\sqrt{4\sigma}$, which favors the $a_0(980)$ and $f_0(980)$ to be  
four-quark candidates. On the other hand, the effective heavy quark  
confinement potential is a direct result of this model.\\
\end{abstract}
\pacs{12.38.Aw, 12.40.Yx} 
\maketitle

Quantum Chromodynamics(QCD) is 
commonly accepted as the correct theory for 
strong interaction. In contrast to its success in the high energy 
regime, there are many questions unsolved in the low energy sector due to
its nonperturbative characteristics. Specifically, confinement is still 
a conjecture and can not be derived from the first principle. Various 
QCD-based models were supposed to describe the hadron spectrum but the 
results are always model dependent.  Apart form the confinement, there are 
other open questions to be answered, such as why the lightest baryon, 
say, nucleons, are so heavy, since they 
are composed of light quarks, how can many mesons and baryons be arranged 
on Regge trajectories $J=\alpha + \alpha' M^2 $ ( $J$ 
and $M$ are the angular momentum and the mass of the hadron, 
respectively, and the
parameters $\alpha$ and $\alpha'$ are the 
interception and the Regge
slope)~\cite{regge1,regge2}, how to explain the 
approximate 
constant mass-square difference~\cite{data} of $1 {}^3 S_1$ and $1 {}^1 
S_0$ 
meson states? The aim of this work is to give a tentative interpretations 
of these puzzles with a new model. 
\par
It is a natural assumption that the whole world should be a color singlet, 
since there are no isolated color charges found from experiments. 
Therefore, theoretically any isolated color charge should be considered 
with an implicit condition that there must exist other color charges in 
the same time so that all the charges add up to a singlet. On the other 
hand, QCD vacuum is topologically non-trivial and has non-zero gluon
condensate and quark condensate, which is in contrast with the case of 
QED. As a result of above, when the behavior of a color charge is considered, 
the QCD vacuum can be viewed as a background color field ( 
called {\it colored vacuum} here) interacting with the charge considered, 
due to the existence of other color charges. Or in other words, the 
universal property of this 'color coherence' implies that color charges 
should be considered in '(generalized) hadron systems', but there
are no limits assumed on the sizes of these 'hadrons' at present. As a
model, we choose the hadron rest frame as the preferred reference frame to 
study the motions of color charges in this work, and require that color 
charges obey relativistic kinetics. 
\par
Intuitively and in the classical meaning, vacuum should be spatially 
homogeneous, or in other words, the color electric field ${\bf E}$ and 
the color magnetic field ${\bf B}$ felt 
by color charges in the vacuum should be identical, if the inter-charge 
Coulomb type interaction is ignored at large enough separation. This kind 
of vacuum can be described by an accessory quantity $\hat{\hat{\sigma}}$, 
called vacuum tensor, which is a constant spatial parallel vector and is 
invariant in spatial translation and rotation,
\begin{equation}
\hat{\hat{\sigma}}=\sigma\delta_{ij}\hat{\bf x}_i\hat{\bf x}_j,
\end{equation}
where $\hat{\bf x}_i$ is the unit vector of the spatial direction $i$.
Its most attractive feature is that it has uniform projection in any 
direction. For example, its projection in the direction of 
$\hat{\bf r}=(sin\theta cos\phi, sin\theta sin\phi,cos\theta)$ is
\begin{equation}
{\bf \sigma}(\theta,\phi)\equiv \hat{\hat{\sigma}}\cdot \hat{\bf 
r}=\sigma\hat{\bf r}.
\end{equation}
\par
In a definite frame of reference and a definite reference point, there 
are two definite directions for a color charge moving in the vacuum, say,
the directions of the velocity $\hat{\bf v}$ and the angular momentum of 
the charge $\hat{\bf L}$.
Purely artificially, we define the color electric field strength felt by 
the color charge in the vacuum as $ {\bf E}= \hat{\hat{\sigma}}\cdot 
\hat{\bf v}$ and the magnetic field strength as ${\bf B}= 
\hat{\hat{\sigma}}\cdot\hat{\bf L}$. In analogy with electromagnetic 
theory, the equation of motion of the color charge in the colored vacuum 
is written as 
\begin{equation}
\label{motion}
\frac{d{\bf p}}{dt}=
   -\sigma \frac{{\bf v}}{|{\bf v}|}-\sigma{\bf v}\times\hat{{\bf 
L}},  
\end{equation}
where ${\bf v}$ is the velocity and the minus signs are chosen to 
guarantee the conservations of energy and angular momentum (see below). 
It is obvious from the equation that angular momentum and the energy of 
the charge vary during the moving in the vacuum. For a single 
quark moving away from a place very close to a specific point $O$, the 
variation of angular momentum referring 
to $O$ at time $\tau$ is 
\begin{eqnarray}
\Delta L&=&\int\limits_{0}^{\tau}{\bf r}\times \frac{d{\bf 
p(t)}}{dt}dt\\\nonumber
&=&-\frac{1}{2}\sigma r(\tau)^2-\sigma\int\limits_{0}^{\tau}dt
{\bf r}(t)\times \frac{{\bf v}(t)}{|{\bf v}(t)|}\\\nonumber
&\equiv& -\frac{1}{2}\sigma R^2 - \sigma L_2,  
\end{eqnarray}
and the energy variation is 
\begin{equation}
\Delta E =-\sigma\int\limits_{0}^{\tau}vdt\equiv -\sigma \Delta S.
\end{equation}
Given the initial velocity $v_0$ and the final velocity $v\sim 0$
(in the practical calculation, the final velocity is set to be $v_f=0.001$), 
the Eqn.~\ref{motion} can be solved numerically with the results shown 
in the table. 
\begin{table}[t]
\label{tab:1}
\begin{center}
\begin{tabular}{cccc}\hline\hline
$v_0$   & $(1-v_0^2)^{-\frac{1}{2}}$ & $\Delta S/R$ & $R/\sqrt{L_2}$\\ 
\hline
0.5     &       1.155	       &     1.008    &    1.426-1.47  \\
0.9     &       2.294	       &     1.059    &      1.377     \\ 
0.99    &       7.089          &     1.202    &      1.420     \\ 
0.999   &       22.37          &     1.310    &      1.415     \\ 
0.9999  &       70.71          &     1.390    &	     1.410     \\ 
0.99999 &       223.6          &     1.411    &      1.422     \\ 
0.999999&       707.1          &     1.414    &      1.415     \\ 
$\sim 1$&   $ \sim\infty$      &$\sim\sqrt{2}$&	$\sim\sqrt{2}$ \\ 
\hline\hline 
\end{tabular} 
\caption{
The numerical solution of Eqn.\ref{motion}. The initial energy 
of a massive color charge is proportional to $(1-v_0^2)^{-\frac{1}{2}}$.}
\end{center}
\end{table}

\par
Table~\ref{tab:1} implies a very interesting relation for an 
ultrarelativistic color charge moving in the QCD vacuum
\begin{equation}
\label{ratio}
\frac{\Delta S}{R}=\sqrt{2}~~~~for~~~~~ v_0\rightarrow 1.
\end{equation}
\par
Now we apply the above logic to a $q\bar{q}$ meson state (in the paper $q$ 
denotes the a light quark, while $Q$ denotes a heavy quark). 
In the meson's rest frame, imagining the constituent quark-antiquark pair 
is excited with initial energy and momentum $(E,\pm {\bf p})$ and 
the initial angular momentum ${\bf J}$ referring to the center of mass. 
According to the equation of motion, they will exchange energy with the 
vacuum during their flight and will finally be quasi-static, thus the meson 
is made up of the quasi-static quark pair plus the excited vacuum
(after absorbing the energy and angular momentum from the quarks). 
With the fact that the mass of light quark is $m_q\sim 5 MeV$ and the 
mass of a typical light meson is $M\sim 1-2GeV$, the initial 
velocity of a light quark is $v_0>0.99995$ and the relation Eq.\ref{ratio} 
is approximately correct. If we assume that the initial position of the 
quark pair is very near the 
center of mass and the final separation is $2R$, the mass $M$ of the meson is 
\begin{equation}
M=2E=2m_q+2\sigma\Delta S\approx 2\sqrt{2}\sigma R,
\end{equation}     
while the total angular momentum $J$ is 
\begin{equation}
J=2|\Delta L|=\sigma R^2+2\sigma(\frac{R}{\sqrt{2}})^2=2\sigma R^2.
\end{equation}
$M$ and $J$ satisfy the relation
\begin{equation}
J=\frac{1}{4\sigma}M^2.
\end{equation}
\par
It is encouraging that we can obtain the Regge relation even in so simple 
a classic model. However, the equation of motion implies that the 
constituents be static finally which is of course not the fact in the real 
world. To give a more delicate description of hadron in quantum theory, 
we put forward the following assumptions:
\par i) Hadrons are strongly coupled system made up of constituents (such 
as quarks and gluons) and the excited color vacuum.
\par ii) Constituents reside in stationary states in hadrons.
\par iii) As far as a constituent is concerned, other constituent(s) and 
the vacuum act effectively as a background field which can be described by 
the parallel vector defined above.
\par iv) Since constituents are in stationary states, the electric-type
interactions in Eqn.\ref{motion} can be omitted. Thus constituents in 
hadrons are analogous to electrons moving in a identical magnetism.
\par
We assume the stationary equation of a constituent in a hadron is 
\begin{equation}
E^2\phi = (m^2 + ({\bf p} + {\bf A})^2 )\phi,
\end{equation}
where $\phi$ is the state function, $m$ is the mass of the constituent, 
and $E$ is the quasi-energy ( see below ). ${\bf A}$ is the 'vector 
potential' due to the color magnetic field of the vacuum. If we choose 
the direction of the angular momentum of a hadron ( also 
the direction of the color magnetic field ) as $z$-axis, the center 
of mass as the reference point, and write the vector potential as ${\bf A} 
= \frac{1}{2}{\bf B}\times {\bf r}$, we have, 
\begin{equation}
\label{dirac3}
(E^2-m^2)\phi=[{\bf p}^2+\frac{1}{4}\sigma^2 (x^2+y^2)-\sigma L_z]\phi,
\end{equation} 
where $L_z = xp_y-yp_x$. The eigenvalues can be directly read out as 
\begin{equation}
E_n (p_z)^2=m^2+p_z^2+\sigma(2n+1).
\end{equation}
\par
We call $E_n$ the quasi-energy because it is meaningless to say 
the energy of a constituent in a strongly coupled system, say, a hadron. 
Instead we introduce a phenomenological mass formula for a multi-quark 
system ( in the hadron rest frame )
\begin{equation}
\label{mass}
M^2=\sum\limits_i{(E_n^i)^2}=\sum\limits_i m_i^2 +\sigma\sum\limits_i 
(2n_i+1),
\end{equation}
which can describe many properties of hadron spectra as follows.
\par{\bf Mesons} For a meson composed of two light constituent quarks in 
its rest frame, the average momenta of the two constituents should be equal, 
so that $n_1$ and $n_2$ in Eqn.\ref{mass} take the same value $n$ and the 
mass formula becomes
\begin{equation}
M^2=\Delta^2(m_1,m_2)+2(2n+1)\sigma,
\end{equation}
where $\Delta^2(m_1,m_2)$ is the mass term comes from other mechanisms, 
such as chiral symmetry breaking. We argue that $\Delta(m_1,m_2)$ is equal 
to the mass of pseudoscalar counterpart, $M_{0^{-}}$, and rewrite the 
above formula as
\begin{equation}
M^2 = M_{0^-}^2+2(2n+1)\sigma.
\end{equation}
The second term is obviously the vacuum effect and comes from the vacuum 
excitations. To comply with the classic scenario discussed above, we 
argue additionally that $n$ is the angular momentum carried by the vacuum
excitations and is denoted with $J_v$ from now on. Maybe we can go further 
to assume that the quantum of the vacuum excitation is with $J^{PC} = 
1^{++}$ and contributes $4\sigma$ to the $M^2$.  The main results of 
this mass formula is:
\par
i) Pseudoscalars, such as $\pi, K$, etc, are Goldstone particles due to
the chiral symmetry breaking, and are excluded in the scenario of this 
work.
\par 
ii) The mass-square differences of the lowest $q\bar 
q$-mesons ($J_v=0$) in this model, say, $\rho, K^*$, etc, and their 
psuedoscalar counterparts are $ M^2 - M_{0^-}^2 = 2\sigma $, which is 
flavor independent. This gives a natural explanation of the phenomenon that
experimentally the mass-square differences of ${}^3 S_1$ states and the 
corresponding ${}^1 S_0$ states are approximately constant, namely, $0.55 
GeV^2$, for $q\bar{q}$ and $q\bar{Q}$ mesons. This is not applied to 
$Q\bar{Q}$ system. A possible reason is that the inter-quark 
Coulomb interaction plays an important role in $Q\bar{Q}$ system, since 
the 
Coulomb effects is proportional to the reduced mass $\mu=m_Q/2$. While 
in the case of $q\bar{q}$ and $q\bar{Q}$, the reduced mass is $\mu\sim 
m_q \sim 0$.
\par
iii) Masses of mesons made up of light constituents are dominated by the 
vacuum effects, especially for large $J_v$. The experimental implication 
of this fact is that light mesons (or more 
generally, hadrons) can be sorted into various 'mass bands'. Mesons 
in the same band have similar masses which are around the mass value 
$\sqrt{(M_{0^-}^2+2\sigma)+4\sigma J_v}$, and the variations from 
this value are results of other mechanisms, such as Coulomb interaction, 
orbit-spin coupling, etc. In other words, we can   
take this mass formula as a generalized Regge relation,
\begin{equation}
J_v = \alpha + \alpha' M^2,
\end{equation} 
with the Regge slope $\alpha'=1/(4\sigma)$ and the interception 
$\alpha= -M_{0^-}^2-2\sigma$. Using the experimental value 
$\alpha'^{-1}\sim 1.1 GeV^2$ we obtain the 
value of $\sigma\approx 0.28 GeV^2$, which can be also determined from
the mass-square differences between $1{}^3 S_1$ and $ 1{}^1 S_0$ 
meson states, $\Delta M^2 \sim 0.55 GeV^2 = 2\sigma$. 
\par
For simplicity, we choose strange mesons for example to illustrate 
the scenario described above, since there exists many non-$q\bar{q}$ 
candidates in the unflavored meson spectra. 
\par
We maintain the $P$ parity and the $C$ parity assignments $P=(-)^{L-1}$ 
and $C=(-)^{L+S}$ for mesons. $J_v = 0$ state corresponds to a mass 
$M=0.891 GeV$, complies with $K^{\ast}(1^-)(892)$. $J_v=1$ corresponds to 
a mass $M=1.387 
GeV$. There are four strange mesons around this mass value, such as 
$K_1(1^+)(1400)$, $K_0^\ast(0^+)(1430)$, $K^\ast(1^-)(1410)$, and 
$K_2^\ast (2^+)(1430)$. ${K_2^\ast, K_1, K_0^\ast}$ may be sorted into a 
$(J_v=1,L=1)$ triplet, $K^\ast$ might be a state of $(J_v=1, L=0)$. 
$J_v=2$ corresponds to a mass $M=1.744 GeV$. There are five strange mesons 
in this range, which are $K^{\ast}(1^-)(1680)$, $K_2(2^-)(1770)$,
$K_3^\ast(3^-)(1780)$, 
$K_2(2^-)(1820)$, and $K(0^-)(1830)$. These states might be sorted into a 
multiplet $(J_v=2, L=0$, or $2)$. $J_v=3$ corresponds to a mass $M=2.040 
GeV$. There are three strange mesons in this range, which are 
$K_4^\ast (4^+)(2.045)$, $K_s^\ast (2^+)(1980)$ and $K_0^\ast(0^+)(1950)$. 
The first two mesons might be the candidates for $(J_v=3, L=1)$ triplet, 
while the third might be $(J_v=3, L=3)$. The strange meson $K(3^+)(2320)$ 
might be a state of $(J_v=4, L=1)$.
\par
{\bf Baryons} Similarly, the mass formula of baryons can be written as 
\begin{equation}
M_b=\sqrt{\delta(m) + (2n_1+2n_2+2n_3+3)\sigma}.
\end{equation}
If the constituents are all light quarks, $\delta$ is small and can be 
ignored temporarily, the lowest mass of a isospin-$\frac{1}{2}$ baryon is 
approximately 
$\sqrt{3\sigma}=0.918 GeV$, which implies that nucleons are candidate 
states of $(n_1=0, n_2=0, n_3=0)$. The first excited state has the mass
$M=\sqrt{7\sigma}=1.4 GeV$ and favors the Roper to be its candidate. For 
the isospin-$\frac{3}{2}$ $\Delta^{++}(1232)$, the three constituents 
reside in 
the same isospin state $|II_3\rangle =|\frac{1}{2}\frac{1}{2}\rangle$. 
Taking into account that the spin of 
each constituent have two possible values $s_z=\frac{1}{2},-\frac{1}{2}$, 
Pauli's exclusion principle requires that $n_1, n_2, n_3$ can not take 
the same value simultaneously. Thus the lowest state for isospin-$3/2$ 
baryon should be $(n_1,n_2,n_3)=(1,0,0),(0,1,0)$, or $(0,0,1)$ with the mass 
$M_b=\sqrt{5\sigma}=1.18GeV$.
\par
It is remarkable that the color degree of freedom was introduced
originally to solve the spin-statistics 'puzzle'
of $\Delta^{++}$~\cite{nambu} and hadrons are thus conjectured to be 
color singlets. However in our logic, this is not a conjecture but a
consequence of the following deduction: since the universe is a color 
singlet, a non-singlet system must be accompanied by other non-singlet
system(s), and all the systems make up a bound state with energy described  
by Eqn.\ \ref{mass}, which increases with the separations between these 
systems (see below). By the way, the color index of a quark always 
changes after interacting with gauge field and may not be taken as 
a good quantum number of a quantum state. As far as this is concerned, our 
explanation of the $\Delta^{++}$ spin-statistics 'puzzle' seems more 
reasonable. This interpretation also implies that the spin of 
$\Delta$ might not be contributed totally by quark spins.
\par
From the discussion above, we can get two important mass ratios 
$M_N/M_\rho \approx \sqrt {3/2} \approx 1.22 $ and $M_\Delta/M_N = 
\sqrt{5/3} = 1.29$ (where $M_N$,$M_\rho$, and $M_\Delta$ are the masses of 
nucleon,$\rho$ meson, and $\Delta$ baryon, respectively), which are in 
very 
good agreement with the experiment values $ 1.22$ and $1.31$.  
\par If our logic above is correct, the lowest mass of four quark state 
$qq\bar{q}\bar{q}$ can be  predicted to be $M\approx \sqrt{4\sigma}=1.060 
GeV$. Experimentally, scalar mesons  $a_0(980)$ and 
$f_0(980)$ lie in this range and have been suggested to be candidates 
of four-quark state. 
\par
{\bf Confinement} Confinement is also a directly result of our model. 
Taking $q\bar{q}$ system for instance, if a constituent resides in the 
stationary state $|n\rangle$, its average distance from the mass center, 
or 'cyclotron radius', can be derived semi-classically as~\cite{radius} 
\begin{equation}
r=\sqrt{\langle x^2+y^2\rangle}=\sqrt{\frac{(2n+1)}{\sigma}}.
\end{equation}
The relation between the energy of the system and the average separation 
$d$ of the two light quarks is 
\begin{equation}
M=\sqrt{M_{0^-}^2+2(2n+1)\sigma}\approx \sigma \sqrt{2}\sigma r \equiv 
\frac{\sqrt{2}}{2}\sigma d,
\end{equation}
which means the energy of the system is proportional to the distance of 
two constituents, as is the exact meaning of confinement. 
\par
For heavy quarkonium, we assume that the 
quarks are so heavy that they are almost decoupled from the vacuum gauge 
field and write the energy of the system as
\begin{eqnarray}
E&=&2M_Q+\sqrt{2(2n+1)\sigma}=2M_Q+\sqrt{2}\sigma r\\\nonumber
&\equiv& 2M_Q+\sigma_{Q\bar{Q}} d,
\end{eqnarray}
where $\sigma_{Q\bar{Q}}=\sqrt{2}/2\sigma\approx 0.2 GeV^2$, the so-called 
string 
tension, is in agreement with the commonly used value. The three quark 
potential can be similarly derived,
\begin{equation}
E=3M_Q+V_{3Q}=3M_Q + \sigma \sqrt{r_1^2 +r_2^2+r_3^2}.
\end{equation}
If we denote $L_{min} = (r_1+r_2+r_3)$ and write $V_{3Q}$ as 
$\sigma_{3Q}L_{min}$ (in flux tube model, $L_{min}$ is the minimal value 
of the total length of Y-type color flux tubes linking three quarks), we 
have, 
\begin{equation}
\sqrt{\frac{2}{3}}\approx 0.82\le \frac{\sigma_{3Q}}{\sigma_{Q\bar{Q}}}
\le 1,
\end{equation}
which is exactly the result of lattice QCD where $\sigma_{3Q}\sim 0.9 
\sigma_{Q\bar{Q}}$~\cite{3q} (the authors of {}~\cite{3q} argue that 
$\sigma_{3Q}\approx \sigma_{Q\bar{Q}}$ be a universal relation).
\par
In summary, starting from a simple classical scenario, we developed a 
model for hadron systems. A hadron can be viewed as a strongly coupled 
system of constituents and the vacuum background field which can be 
described by a parallel vector. In this model, the Regge relation and the 
confinement can be neatly derived, and the hadron mass spectra can be 
consistently described by a simple mass formula. The most interesting
result of this model is that the the Regge slope, the string tension
parameter in the effective heavy-quark potential, and the mass-square 
differences between $1{}^3 S_1$ and $ 1{}^1 S_0$ meson states, can be 
represented by a single constant, $\sigma \approx 0.28 GeV^2$, which 
is the amplitude of the colored vacuum background field.   
 
\par 
 This work is supported by the Natural Science Foundation of China under 
the Grant No. 10075051, No. 12035040, and the CAS Knowledge Innovation 
Project(No. KJCX2-SW-N02). The author thanks Prof. J.-M. Wu and Prof. Z. 
Chang of IHEP for valuable discussions. The author also thanks the 
hospitality of ICTP when this work was prepared and Prof. K. F. Liu of 
University of Kentucky when the manuscript was revised.

\end{document}